\documentclass[%
 reprint,
superscriptaddress,
 amsmath,amssymb,
 prl,
]{revtex4-1}

\usepackage{graphicx}
\usepackage{epstopdf}

\usepackage{dcolumn}
\usepackage{bm}
\usepackage{hyperref}

\begin{document}

\preprint{APS/123-QED}

\title{Competition between Spin Echo and Spin Self-Rephasing \\in a Trapped Atom Interferometer}

\author{C. Solaro}
\author{A. Bonnin}
\affiliation{SYRTE, Observatoire de Paris, PSL Research University, CNRS, Sorbonne Universit\'es, \\ UPMC Universit\'e Paris 06, LNE, 61 Avenue de l'Observatoire, 75014 Paris, France}
\author{F. Combes}
\affiliation{Laboratoire de Physique des Solides, CNRS UMR 8502, Universit\'e Paris-Sud, Universit\'e Paris-Saclay, F-91405 Orsay Cedex, France}
\author{M. Lopez}
\author{X. Alauze}
\affiliation{SYRTE, Observatoire de Paris, PSL Research University, CNRS, Sorbonne Universit\'es, \\ UPMC Universit\'e Paris 06, LNE, 61 Avenue de l'Observatoire, 75014 Paris, France}
\author{J.-N. Fuchs}
\affiliation{Laboratoire de Physique des Solides, CNRS UMR 8502, Universit\'e Paris-Sud, Universit\'e Paris-Saclay, F-91405 Orsay Cedex, France}
\affiliation{Laboratoire de Physique Th\' eorique de la Mati\` ere Condens\' ee, CNRS UMR 7600, \\ Universit\'e Pierre et Marie Curie, 4 place Jussieu, 75252 Paris Cedex 05, France}
\author{F. Pi\'{e}chon}
\affiliation{Laboratoire de Physique des Solides, CNRS UMR 8502, Universit\'e Paris-Sud, Universit\'e Paris-Saclay, F-91405 Orsay Cedex, France}
\author{F. Pereira dos Santos}
\affiliation{SYRTE, Observatoire de Paris, PSL Research University, CNRS, Sorbonne Universit\'es, \\ UPMC Universit\'e Paris 06, LNE, 61 Avenue de l'Observatoire, 75014 Paris, France}

\date{\today}

\begin{abstract}
We perform Ramsey interferometry on an ultracold $^{87}$Rb ensemble confined in an optical dipole trap. We use a $\pi$-pulse set at the middle of the interferometer to restore the coherence of the spin ensemble by canceling out phase inhomogeneities and creating a spin echo in the contrast. However, for high atomic densities, we observe the opposite behavior: the $\pi$-pulse accelerates the dephasing of the spin ensemble leading to a faster contrast decay of the interferometer. We understand this phenomenon as a competition between the spin-echo technique and an exchange-interaction driven spin self-rephasing mechanism based on the identical spin rotation effect. Our experimental data is well reproduced by a numerical model.
\begin{description}
\item[PACS numbers] 05.30.-d, 37.25.+k, 32.80.Qk, 67.85.-d
\end{description}
\end{abstract}
\pacs{05.30.-d}\pacs{37.25.+k}\pacs{32.80.Qk}\pacs{67.85.-d}

\maketitle

A long coherence time is crucial for the coherent manipulation of quantum systems. In quantum information, high-precision spectroscopy as well as in atom interferometry, preventing pure quantum states from decaying into a statistical mixture is challenging. In particular, when manipulating trapped ensembles, particles experience different trapping potentials and their spins precess at different rates creating a deleterious dephasing of coherences. Several techniques have been developed to overcome this limitation and to extend the coherence of the spin ensemble. The use of a magic wavelength for optically trapped atomic clouds \cite{Katori_&al_2003,Katori_&al_2009}, the addition of a compensating field \cite{Kaplan_&al_2002} or the mutual compensation scheme in magnetically trapped Rb ensembles \cite{Lewandowski_&al_2002} have been demonstrated. All such techniques, however, only reduce the dephasing that is slowed down but never canceled. 

A widespread technique that cancels out inhomogeneous dephasing is to create a spin echo via a $\pi$-pulse as originally thought of for NMR spectroscopy \cite{Hahn2_1950} and later on extended to cold gases \cite{Andersen_&al_2003}. The spin-echo technique reverses the inhomogeneous dephasing of a spin ensemble, which significantly increases the coherence time of the quantum system. Especially, inertial atomic sensors highly benefit from spin-echo techniques as, for symmetric interferometers, it cancels out unwanted clock effects \cite{Kasevich_&al_1991}, enabling unprecedented high sensitivities on the measurement of inertial forces \cite{Dutta_&al_2016,Freier_&al_2015,Gillot_&al_2014,Hu_&al_2013,Sorrentino_&al_2014,Biedermann_&al_2015}. In such configurations, however, dephasing sources originating from atom transverse motion \cite{Louchet-Chauvet_&al_2011,Hilico_&al_2015} can only be tackled by cooling the particles to lower temperatures, suggesting the use of dense ultracold gases. Yet, in such a regime, collisional processes can also lead to inhomogeneous broadening and decoherence of the spin ensemble \cite{Harber_&al_2002}. The spin-echo technique can in principle just as well cancel out mean-field shift inhomogeneities which is expected to restore the coherence of a spin ensemble at low temperatures.

\begin{figure}[ht]
\includegraphics[width=0.48\textwidth]{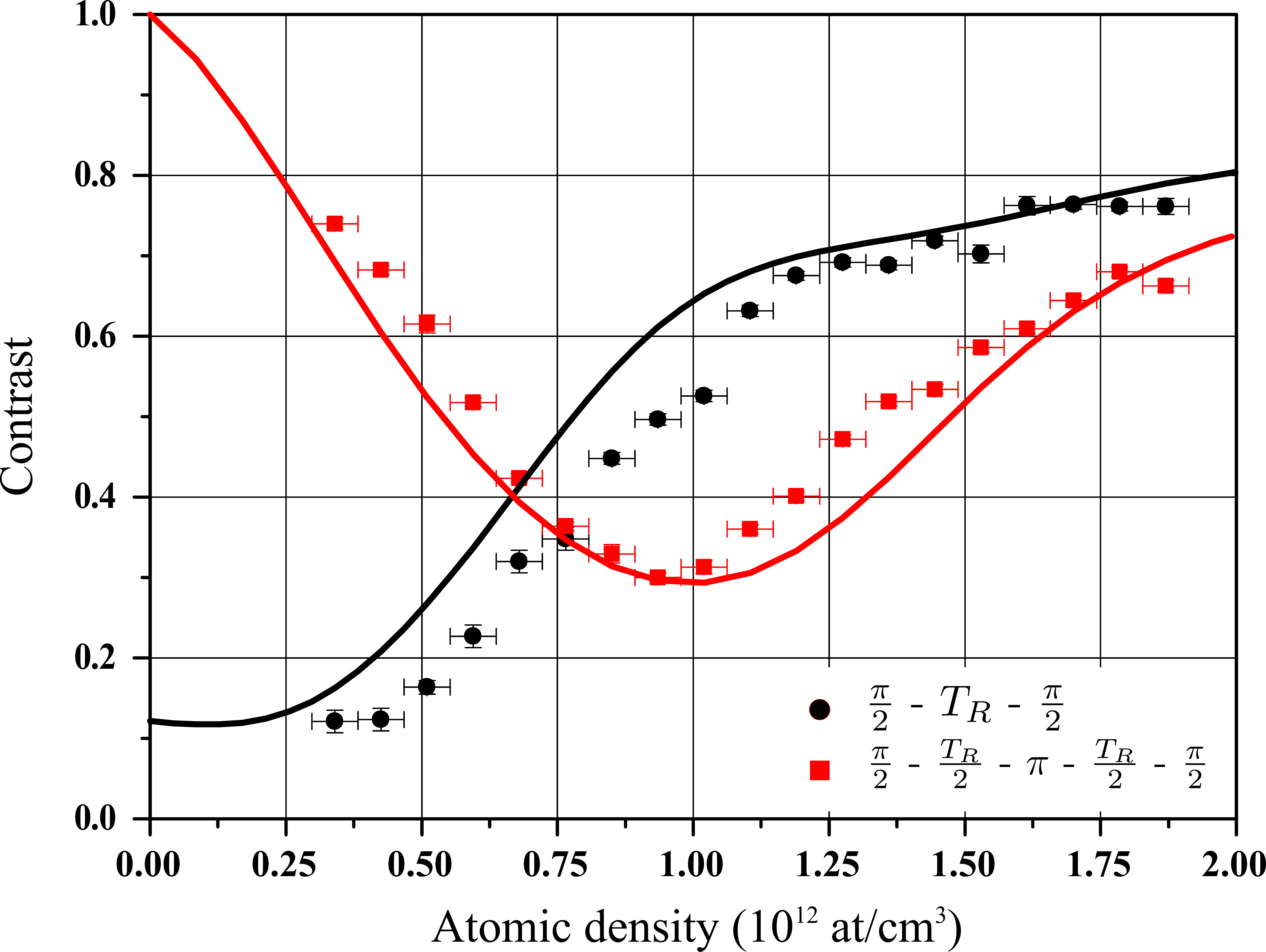}
\caption{(color online). First experiment: Ramsey fringe contrast as a function of the mean atomic density $\bar{n}$ at a fixed Ramsey time $T_R = 0.4$ s. Red squares and black dots correspond to a Ramsey sequence with and without applying a $\pi$-pulse at $t_\pi=T_R/2$ respectively. Lines correspond to our simulation results  \cite{[{See Supplemental Material at [URL will be inserted by publisher] for details regarding our numerical model}] supmat}. Black line and dots: the contrast increases with the atomic density which is characteristic of the spin self-rephasing mechanism \cite{Deutsch_&al_2010}. Red dashed line and squares: the $\pi$-pulse is expected to help canceling out the inhomogeneous broadening leading to a much higher contrast. Surprisingly for densities larger than $0.75\times10^{12}$ at/cm$^{3}$, the $\pi$-pulse leads to a lower contrast which reaches a minimum at an intermediate density $1\times10^{12}$ at/cm$^{3}$. This non-monotonic behavior suggests a competition between spin echo and spin self-rephasing mechanisms.}
\label{Fig1}
\end{figure}

However, using Ramsey interferometry and the spin-echo technique on an ultracold $^{87}$Rb ensemble confined in an optical dipole trap, we observe the opposite behavior: applying a $\pi$-pulse in the middle of the Ramsey sequence increases spin dephasing leading to a faster loss of coherence of the spin ensemble. This unexpected phenomenon results from the interplay between the $\pi$-pulse and a collective spin self-rephasing (SSR) mechanism \cite{Deutsch_&al_2010} due to the cumulative effect of the identical spin rotation effect (ISRE) in a trapped collisionless gas. Originating from atomic interactions and indistinguishability, ISRE is responsible for the exchange mean-field \cite{Lhuillier_Laloe_1982} and is characterized by the exchange rate $\omega_{ex}= 4\pi\hbar a_{12}\bar{n}/m$, where $a_{12}$ is the relevant scattering length \cite{Sortais_These_2001}, $\bar{n}$ the mean atomic density and $m$ the atomic mass. In this Letter, we investigate the interplay between the spin-echo technique and SSR, two different rephasing mechanisms that, surprisingly, do not collaborate.

\textit{Experimental set-up} - In our system, we manipulate laser-cooled ultracold $^{87}$Rb atoms trapped in a 3D optical dipole trap red-detuned far from resonance ($\lambda=1070$ nm). The trap consists of two intersecting beams of 30 and 176 $\mu m$ waists with maximum powers of 6 and 45 W respectively. After loading the trap from a magneto optical trap (MOT), we perform 2 s evaporative cooling to reach a cloud temperature of $T\sim 500$ nK with trap frequencies $\{\omega_1,\omega_2,\omega_3\} = 2\pi\times \{27, 279, 269\}$ Hz. With $\bar{n}$ in the range of $10^{12}$ at/cm$^{3}$, we explore the non-degenerate and collisionless (Knudsen) regime where the trap frequencies are much larger than the rate $\gamma_c$ of lateral collisions. These collisions are velocity changing elastic collisions, also known to be responsible for collisional narrowing \cite{Sagi_&al_2010}. The rate $\gamma_c$ is given by $\sim 4\pi a^2 v_T \bar{n}$, where $a$ is the scattering length \cite{[{This expression for $\gamma_c$ assumes $a_{11} \approx a_{22} \approx a_{12} \approx 100a_0 \approx a$ as for $^{87}$Rb}] gammac} and $v_T=\sqrt{k_BT/m}$ the thermal velocity of the atoms. $\gamma_c \sim 2.4\bar{n}$ s$^{-1}$ with $\bar{n}$ in $10^{12}$ at/cm$^{3}$ which, in our range of densities, remains lower than 5 s$^{-1}$. In order to maximize the contrast of our interferometers, while preserving highest atomic densities, the atoms are carefully polarized into the state $|5s ^2S_{1/2}, F = 1,m_F = 0\rangle$ to avoid spin relaxation \cite{Widera_&al_2005} and to minimize their sensitivity to parasitic magnetic fields. During the early stage of evaporation, after switching on our quantification axis, a $\sigma$-polarized resonant pulse on the $|5s ^2S_{1/2}, F = 1\rangle$ to $|5s ^2P_{3/2}, F'=0\rangle$ transition pumps 70\% of the atoms into the dark state $|F = 1,m_F = 0\rangle$. A combination of microwave and optical pulses is then used after the evaporation to purify the atomic sample leading to a highly polarized spin ensemble where 98$\pm$1$\%$ of the atoms are in the $| F = 1, m_F=0 \rangle$ state. After this preparation sequence, we use resonant microwave field on the $|F = 1,m_F = 0\rangle$ to $|F=2,m_F = 0\rangle$ transition and perform Ramsey interferometry. We can implement a standard Ramsey interferometer \mbox{( $\pi$/2 - $T_R$ - $\pi$/2 )} or add a $\pi$-pulse in the interferometric sequence \mbox{( $\pi$/2 - $t_\pi$ - $\pi$ - $t_2$ - $\pi$/2 )} with $T_R=t_\pi+t_2$. In particular we can realize a symmetric Ramsey interferometer when $t_2=t_\pi$. After the interferometer, the atoms are released from the trap and the populations in the two hyperfine states are measured. This state selective detection is based on fluorescence in horizontal light sheets at the bottom of the vacuum chamber \cite{Clairon_&al_1995}. Since this detection system does not resolve the different $m_F$ states, $|F=2,m_F=\pm1\rangle$ states that are created during the interferometer sequence by spin relaxation contribute as a background, which reduces the contrast by about 15 \% for $T_R=1$ s and $\bar{n}=2\times10^{12}$ at/cm$^{3}$.

\textit{First Experiment} - To illustrate the effect of the spin-echo technique and SSR onto the coherence of the spin ensemble, we display in figure \ref{Fig1} measurements of the Ramsey contrast with and without a $\pi$-pulse for different atomic densities $\bar{n}$. To vary $\bar{n}$, we vary the number of atoms by changing the MOT loading time, accessing densities from 0.3 to $2.5\times10^{12}$ at/cm$^{3}$ which modulate $\omega_{ex}/2\pi=7.5\bar{n}$ Hz with $\bar{n}$ in $10^{12}$ at/cm$^{3}$. The cloud temperature was verified to remain constant within 15\%. In this regime, inhomogeneous dephasing originates both from differential light shift induced by the trapping lasers and from mean-field interactions, and has for characteristic inhomogeneity \mbox{$\Delta_0/2\pi = k_BT/2h\times\delta\alpha/\alpha + 2\sqrt{2}\hbar(a_{11}-a_{22})\bar{n}/m$} using the same definition of $\Delta_0$ as in \cite{Deutsch_&al_2010,Kleine_&al_2011}. Here $\delta\alpha$ and $\alpha$ are the differential and the total light shift per intensity. With $\delta\alpha/\alpha=5.93\times10^{-5}$ and $a_{11}$, $a_{22}$ the relevant scattering lengths \cite{[{We use $a_{11} = 101.284 a_0$, $a_{22} = 94.946 a_0$ and $a_{12} = 99.427 a_0$ as calculated by C. Williams and found in }] Sortais_These_2001} we have $\Delta_0/2\pi \approx 0.3 + 0.7 \bar{n}$ Hz with $\bar{n}$ in $10^{12}$ at/cm$^{3}$. The Ramsey time is $T_R = 2t_\pi = 0.4$ s and the Ramsey fringe contrast is deduced by scanning the phase of the exciting field. Without any $\pi$-pulse, the contrast increases with the atomic density (figure \ref{Fig1} black dots), revealing the efficiency of the SSR mechanism. Applying a $\pi$-pulse is expected to help canceling out the inhomogeneous broadening, thus leading to a much higher contrast. For densities between 0.3 and 0.75$\times10^{12}$ at/cm$^{3}$ the $\pi$-pulse indeed increases the contrast with respect to a standard Ramsey interferometer, but its efficiency is diminished as the density increases (figure \ref{Fig1} red squares). In fact, for higher densities the $\pi$-pulse actually accelerates the dephasing, but in a non-monotonic manner so that the contrast reaches a minimum at an intermediate density 1$\times10^{12}$ at/cm$^{3}$ and then substantially recovers at highest densities. This behavior traduces an unexpected competition between the individual spin echo and the collective SSR mechanisms.

\begin{figure}[!ht]
\includegraphics[width=0.48\textwidth]{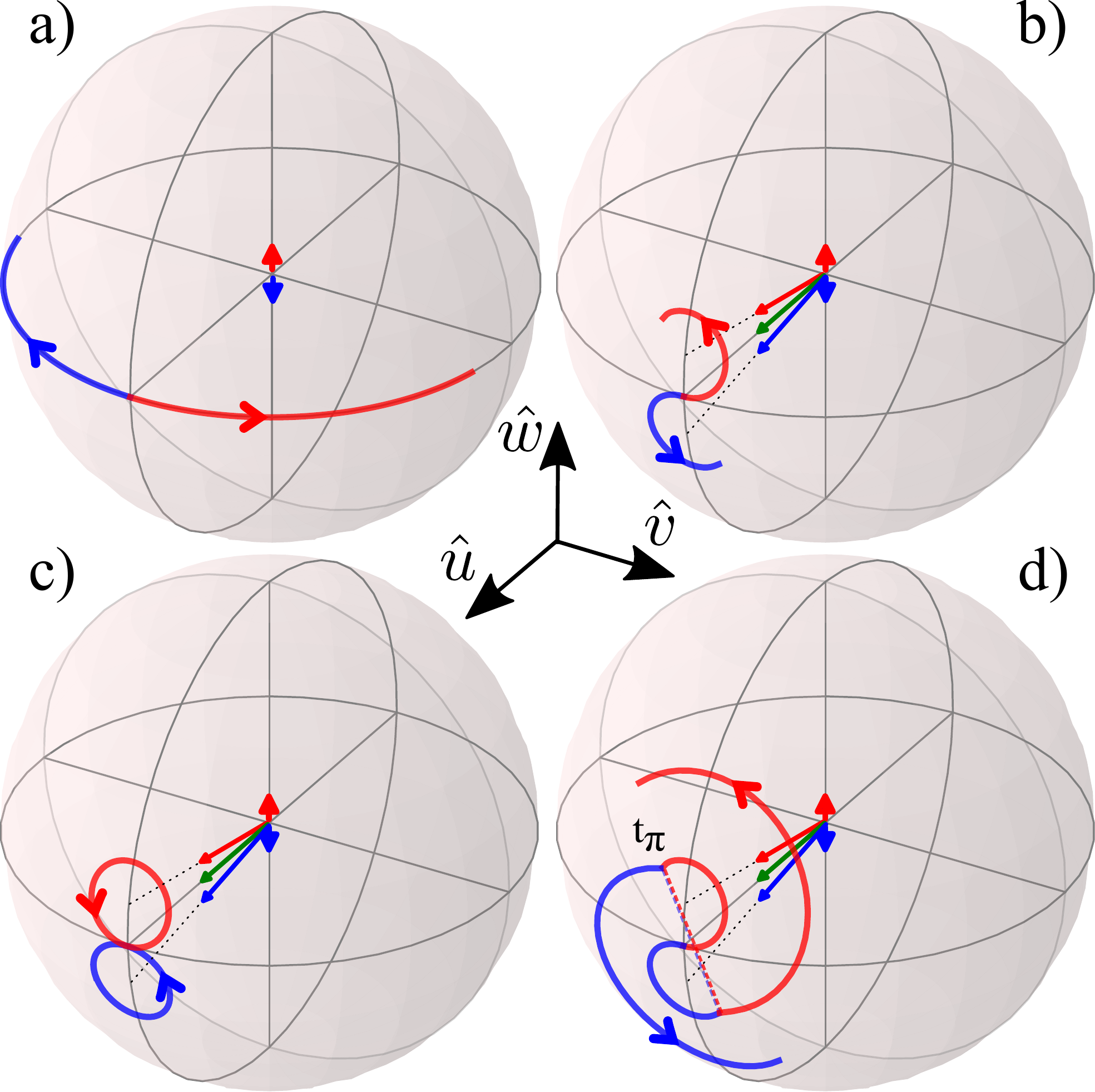}
\caption{(color online). The atomic population is divided into two equal classes of hot (red) and cold (blue) atoms that are represented by their macrospins trajectories on the Bloch sphere. (a) Inhomogeneous dephasing acts as a torque pointing in the $\bm{\hat{w}}$ direction that is of opposite sign for the two classes of atoms (red and blue short arrows). (b) With the ISRE, the effective magnetic field seen by the atoms is the sum of the inhomogeneity and the exchange mean-field proportional to the total spin (green arrow). As a consequence, the hot (cold) macrospin precesses around the red (blue) long arrow, so that if no $\pi$-pulse is applied they rephase at time $T_{ex}$: this is the SSR (c). If one applies a $\pi$-pulse when the two macrospins are out of the equatorial plane $(u,v)$, the rephasing is degraded (d).}
\label{Fig2}
\end{figure}

\begin{figure*}[!ht]
\includegraphics[width=1\textwidth]{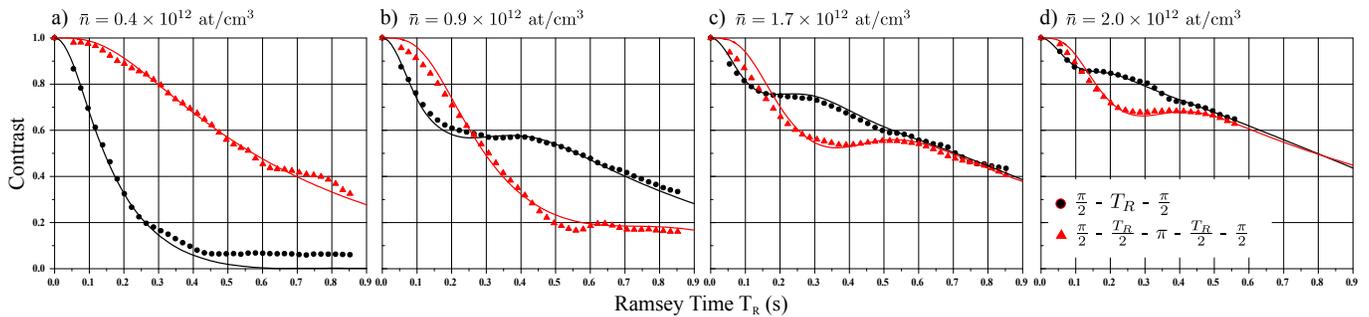}
\caption{(color online) Second experiment: Ramsey contrast versus Ramsey time $T_R$ for standard Ramsey interferometers (black dots) and symmetric Ramsey interferometers with $t_\pi=T_R/2$ (red triangles) for different atomic densities. Lines correspond to numerical simulations. For details regarding our model and the values of fit constants see \cite{supmat}.}
\label{Fig3}
\end{figure*}

\textit{Two macrospins model} - SSR was first observed in a magnetically trapped $^{87}$Rb ensemble \cite{Deutsch_&al_2010} and subsequently employed in an optical trap \cite{Kleine_&al_2011} to extend the coherence time of a spin ensemble. This noteworthy collective mechanism extends the coherence time up to several seconds and leads to revivals of the contrast at periods of the ISRE: $T_{ex}=2\pi/\omega_{ex}$. Following the two macrospins model of \cite{Piechon_&al_2009}, we divide the atomic population into two groups of equal size, each characterized by a macrospin and represent them by their trajectory on the Bloch sphere \cite{Gibble_2010,Deutsch_&al_2010}. The fast (slow) macrospin corresponds to the hot (cold) atoms which occupy higher (lower) transverse states of the 3D harmonic oscillator. For the sake of simplicity, we consider the dynamics in the rotating frame where the total spin always points in the $\bm{\hat{u}}$ direction of the Bloch sphere, and the $\pi$-pulses are rotations of $\pi$ around this axis. Without the ISRE, the effective magnetic field seen by the macrospins only consists in the inhomogeneity which points along $\bm{\hat{w}}$ and in opposite directions for the two macrospins. Hence the fast macrospin rotates towards the right while the slow macrospin rotates towards the left (figure \ref{Fig2}(a)). Applying a $\pi$-pulse at time $t_\pi$ would swap the two macrospins, so that at time 2$t_\pi$ the macrospins would resynchronize leading to a spin echo (not shown). With the ISRE, the effective magnetic field is now the sum of the inhomogeneity and the exchange mean-field which is proportional to the total spin. As a consequence the fast macrospin rotates around this effective magnetic field displayed as the long red arrow in figure \ref{Fig2}(b), staying in the upper hemisphere of the Bloch sphere, while the slow macrospin evolves in the lower hemisphere. If no $\pi$-pulse is applied, the macrospins reach the equatorial plane again in their initial direction $\bm{\hat{u}}$ at the exchange period: the synchronization is perfect (figure \ref{Fig2}(c)). However, if a $\pi$-pulse is applied, swapping the two macrospins positions, their trajectories are not confined to the upper (lower) hemisphere anymore (figure \ref{Fig2}(d)), so that when they reach the equatorial plane again, they are not aligned: synchronization still occurs but is not perfect anymore. The effect of the $\pi$-pulse on SSR depends on the ratio between $t_\pi$ and $T_{ex}$: when the macrospins are back in the equatorial plane the $\pi$-pulse has no effect ($t_\pi=pT_{ex}$ with $p \in \mathbb{N} $), while the effect is worst when the macrospins are maximally out of the equatorial plane ($t_\pi=(p+\frac{1}{2})T_{ex}$). 

\textit{Second Experiment} - To further investigate the competition, we performed standard and symmetric Ramsey interferometers scanning both the phase and the Ramsey time $T_R$ from 0 to 900 ms. We extract the fringe contrast as a function of time for different atomic densities $\bar{n} = \{0.4,0.9,1.7,2\}\times10^{12}$  at/cm$^{3}$ (figure \ref{Fig3}). Black circles correspond to standard Ramsey interferometers and are similar to the ones in \cite{Deutsch_&al_2010} whereas the red triangles correspond to symmetric Ramsey interferometers ($t_2=t_\pi=T_R/2$). For the lowest density, inhomogeneous broadening results in a $1/e$ dephasing time of $T_2^ \star\approx 0.2$ s corresponding to an inhomogeneity of $\Delta_0 \approx 1/T_2^\star \approx 2\pi \times 0.8$ Hz using the same notations as \cite{Kuhr_&al_2005}, and agrees within 15\% of our estimated inhomogeneity. In this regime we verify that the spin-echo technique is very efficient resulting in a $1/e$ coherence time of $T_2' = 0.8$ s (figure \ref{Fig3}(a)). For the highest density, one observes a revival of the standard Ramsey interferometer's contrast at 0.25 s that also corresponds to a local minimum of the symmetric one (figure \ref{Fig3}(d)). Such a behavior is what one would expect from the two macrospins model described above. For the standard Ramsey interferometer the spins are fully resynchronized at the exchange period $T_{ex}$ resulting in a revival of the Ramsey contrast at $T_R = 0.25$ s. For the symmetric Ramsey interferometer, for the same Ramsey time, the atoms were swapped by a $\pi$-pulse at $t_\pi = T_R/2 = 0.125$ s (i.e. at half the exchange period) resulting into an acceleration of the dephasing and the lowest contrast of the interferometer. We observe that same phenomenon for different atomic densities $\bar{n} = \{2,1.7,0.9\}\times10^{12}$  at/cm$^{3}$ at different Ramsey time $T_0\approx\{0.25,0.32,0.5\}$ s with about the same scaling $\bar{n}T_0\approx cst$. Notice that no such other modulation is clearly observed at multiples $p\neq1$ of the exchange period as it is expected from our simple model. This can be explained by our relatively high rate of lateral collisions $\gamma_c$ that perturb SSR, resulting in a damping of the contrast revivals.

\begin{figure}[!ht]
\includegraphics[width=0.48\textwidth]{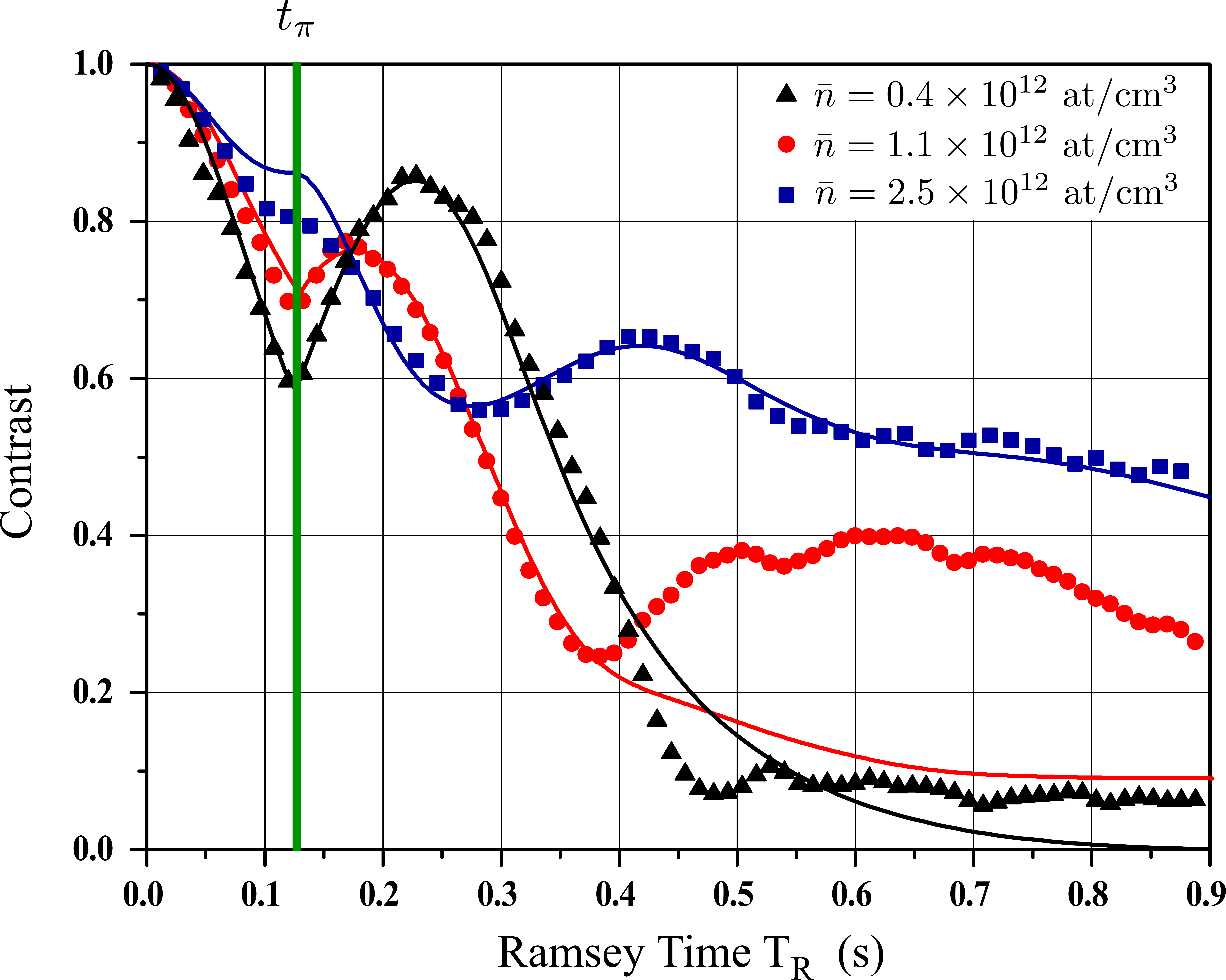}
\caption{(color online) Third experiment: Ramsey contrast versus Ramsey time $T_R$ with a fixed $\pi$-pulse at $t_\pi=125$ ms (green vertical line). Black triangles, red dots and blue squares correspond to the experimental data for different atomic densities. Lines correspond to numerical simulations \cite{supmat}.}
\label{Fig4}
\end{figure}

\textit{Third Experiment} - To further understand the spin dynamics and test our numerical model \cite{supmat}, we measured the evolution of the contrast after a fixed $\pi$-pulse at $t_\pi = 0.125$ s, for three different atomic densities $\bar{n} = \{0.4,1.1,2.5\}\times10^{12}$  at/cm$^{3}$ (figure \ref{Fig4}). We scan the Ramsey time $T_R$ between 0 and 900 ms (with $t_2$ not necessarily equal to $t_\pi$) in order to observe the formation of a spin echo. For the lowest density, we observe a net revival of the contrast at $T_R = 2t_\pi = 0.25$ s. As it is expected when exchange interactions and thus SSR are negligible, the $\pi$-pulse reverses the inhomogeneous dephasing and leads to a spin echo. For $\bar{n}=1.1\times10^{12}$ at/cm$^{3}$, a smaller echo occurs at a time $T_R \approx 0.18$ s $ < 2t_\pi$. This behavior is well reproduced by our model and is a consequence of SSR which tends to faster resynchronize the spins after the $\pi$-pulse. Increasing further the density, one can not see any echo since for such densities the $\pi$-pulse is now mostly a perturbation of SSR. But, for Ramsey times $T_R > 2t_\pi$, SSR (re)occurs at time $T_R \approx t_\pi+T_{ex}$ resulting in a net revival of the contrast at $\approx 0.4$ s. Note that the same behavior is observed for the intermediate atomic density at $T_R \approx 0. 6$ s. It can also be reproduced by the simulation, but with other parameters (in particular a larger exchange rate), and at the expense of a poorer agreement at short Ramsey times. This behaviour remains to be explained.

In summary, we have investigated both experimentally and theoretically the interplay between the spin-echo technique and the spin self-rephasing mechanism in a trapped atom interferometer. In particular, we show that the complex spin dynamics resulting from the competition between these two effects leads to non-trivial evolution of the coherence of the atomic ensemble. We propose a simple two macrospins model to give a qualitative insight into this dynamics. We also find a quantitative agreement between our measurements and the results of a numerical simulation. 

Spin echoes have been used in a recent work \cite{Trotzky_&al_2015} to observe the effect of the ISRE onto the spin diffusion coefficient of a unitary degenerate gas, allowing for the determination of the Leggett-Rice parameter \cite{Leggett_Rice_1968}. Notice that in such a study, by contrast to our situation, SSR is absent as the gas is in the hydrodynamic rather than collisionless regime.

Our results illustrate the crucial role played by atomic interactions, in some case deleterious, in others beneficial, in the dynamics of quantum sensors based on ultracold atoms. In particular, the spin dynamics and coherence of atom interferometers depend on the geometry of the sensor, through the details of the pulse sequence one uses. Whereas the spin-echo technique is not used in standard Ramsey microwave interferometry for clock spectroscopy and frequency measurements, it is widely used in other interferometer configurations as for example for the measurement of inertial forces, where $\pi$-pulses are used to cancel out all clocks effects. However, in such interferometers, the two partial wavepackets associated to the two internal states do not overlap perfectly and ISRE is expected to be weaker. The dependance of SSR with this overlap can be studied in an interferometer using atoms trapped in a lattice, such as \cite{Pelle_&al_2013}. Another interesting perspective would be to investigate if a modified $\pi$-pulse (e.g. a \textit{mirror} pulse \cite{supmat}) would allow producing better spin echoes in the presence of ISRE. Similar issues are currently being studied in the context of many-body interacting quantum systems \cite{Engl_&al_2016}.

\begin{acknowledgments}
We thank Peter Rosenbusch, Wilfried Maineult and Kurt Gibble for useful discussions. We also acknowledge financial support by the IDEX PSL (ANR-10-IDEX-0001-02 PSL) and ANR (ANR-13-BS04-0003-01). A.B. thanks the Labex First-TF for financial support.
\end{acknowledgments}

\bibliographystyle{apsrev4-1}
\bibliography{Bibliographie}

\end{document}


\preprint{APS/123-QED}

\title{Supplemental material to: \\ Competition between Spin Echo and Spin Self-Rephasing \\in a Trapped Atom Interferometer}

\author{C. Solaro}
\author{A. Bonnin}
\affiliation{SYRTE, Observatoire de Paris, PSL Research University, CNRS, Sorbonne Universit\'es, \\ UPMC Universit\'e Paris 06, LNE, 61 Avenue de l'Observatoire, 75014 Paris, France}
\author{F. Combes}
\affiliation{Laboratoire de Physique des Solides, CNRS UMR 8502, Universit\'e Paris-Sud, Universit\'e Paris-Saclay, F-91405 Orsay Cedex, France}
\author{M. Lopez}
\author{X. Alauze}
\affiliation{SYRTE, Observatoire de Paris, PSL Research University, CNRS, Sorbonne Universit\'es, \\ UPMC Universit\'e Paris 06, LNE, 61 Avenue de l'Observatoire, 75014 Paris, France}
\author{J.-N. Fuchs}
\affiliation{Laboratoire de Physique des Solides, CNRS UMR 8502, Universit\'e Paris-Sud, Universit\'e Paris-Saclay, F-91405 Orsay Cedex, France}
\affiliation{Laboratoire de Physique Th\' eorique de la Mati\` ere Condens\' ee, CNRS UMR 7600, \\ Universit\'e Pierre et Marie Curie, 4 place Jussieu, 75252 Paris Cedex 05, France}
\author{F. Pi\'{e}chon}
\affiliation{Laboratoire de Physique des Solides, CNRS UMR 8502, Universit\'e Paris-Sud, Universit\'e Paris-Saclay, F-91405 Orsay Cedex, France}
\author{F. Pereira Dos Santos}
\affiliation{SYRTE, Observatoire de Paris, PSL Research University, CNRS, Sorbonne Universit\'es, \\ UPMC Universit\'e Paris 06, LNE, 61 Avenue de l'Observatoire, 75014 Paris, France}

\date{\today}

\maketitle

\onecolumngrid

\section{Model for spin dynamics simulations} 

\subsection{Spin dynamics in energy space and $\pi$-pulse}

For the quantitative understanding of the interplay between spin self-rephasing and the spin-echo technique, we use the same model as described in \cite{Deutsch_&al_2010}. We here briefly remind the line of reasoning that leads to this model. The trap is approximated by a 3D harmonic oscillator and the motion of atoms in the trap is treated semiclassically. For the considered atom densities and trap geometry, the rate of lateral collisions is small compared to the trap frequencies $\gamma_c \ll \omega_{x,y,z}$. In this situation, the atoms have time to perform many back and forth periodic motion in the trap before any lateral collision takes place. In such a Knudsen regime, averaging over this fast periodic orbital motion enables to regroup the atoms according to their motional energy $E$. The effective spin dynamics of the gas can then be described through coupled Bloch equations of motion for energy dependent spin vectors $\bm{S}(E,t)$ \cite{Deutsch_&al_2010}:
\begin{equation}
\partial_t \bm{S}(E,t)\approx [\Delta_0 E \bm{\hat{w}} + \omega_{ex}\bm{\bar{S}}(t)]\times\bm{S}(E,t)-\gamma_c[\bm{S}(E,t)-\bm{\bar{S}}(t)]
\label{skees}
\end{equation}
with $\bm{\bar{S}}$ being the energy-average spin:
\begin{equation*}
\bm{\bar{S}}(t)=\int_0^\infty dE'\frac{E'^2}{2}e^{-E'}\bm{S}(E',t).
\end{equation*}
and $E\geq 0$ in units of $k_BT$. The first term on the righthand side of (\ref{skees}), represents the inhomogeneous precession (dephasing) which originates both from differential light shift induced by the trapping lasers and from mean-field interactions. It is linear in $E$ for a harmonic oscillator and has for characteristic inhomogeneity $\Delta_0$. The second term represents the exchange mean-field that originates from the identical spin rotation effect (ISRE). It is taken here in the infinite-range approximation. This term is non-linear in spin density and is responsible for the collective spin self-rephasing mechanism \cite{Deutsch_&al_2010}. The last term is the effective spin relaxation due to lateral collisions. The initial state, immediately after the first $\pi/2$-pulse, is taken to be $\bm{S}(E,t=0)=\bm{\hat{u}}$. In absence of the spin relaxation term, the spin vectors $\bm{S}(E,t)$ keep their unit norm at any time. The contrast at time $t$ is given by $C(t)=|\bm{\bar{S}}(t)|$. 

The $\pi$-pulse is implemented by performing the instantaneous transformation $(S_u,S_v,S_w)\rightarrow (S_u,-S_v,-S_w)$ at time $t=t_\pi$, for each $\bm{S}(E,t_\pi)$. After this transformation, the spin dynamics is still determined from the equation of motion (\ref{skees}). Consider first the case where both exchange mean-field and relaxation terms are absent. In this situation, it is well known that the forward evolution after the $\pi$-pulse is equivalent to a backward evolution from the initial state $\bm{S}(E,t=t_\pi)$, such that at time $t=2t_\pi$ one recovers $\bm{S}(E,t=2t_\pi)=\bm{S}(E,t=0)$ leading to a spin echo. In other words the $\pi$-pulse is equivalent to a time reversal operation $t \rightarrow -t$. In the presence of exchange mean-field, the $\pi$-pulse is no more a simple time reversal operation. In fact, within the infinite range approximation, it appears now to be equivalent to the transformation $t,\omega_{ex} \rightarrow -t,-\omega_{ex}$. A further complication is that the exchange mean-field is a non linear spin term. The latter two properties makes it hard to anticipate if and how spin echo and spin self-rephasing mechanisms will cooperate or compete.

Interestingly, we note that in the presence of the exchange mean field it is still possible to find a class of instantaneous transformations equivalent 
to a pure time reversal operation $t \rightarrow -t$.
Each of these transformation consists of a \textit{mirror} operation which reverses the effective spin component perpendicular to a mirror plane that contains the $\bm{\hat{w}}$ axis. 
As an example, for the mirror plane defined by axes $\bm{\hat{u}},\bm{\hat{w}}$ the \textit{mirror} transformation consists in reversing the $v$ spin component: 
$(S_u,S_v,S_w)\rightarrow (S_u,-S_v,S_w)$. Some preliminary simulation results suggest that by performing such a \textit{mirror} transformation 
at a time $t=t_m$ for each $\bm{S}(E,t_m)$, we obtain an almost perfect spin-echo $\bm{S}(E,t=2t_m)=\bm{S}(E,t=0)$ at time $t=2 t_m$, 
despite the non linearity of the exchange mean field. A more systematic study is needed to confirm and widen these preliminary results. 
Nevertheless, we stress that these \textit{mirror} transformations are not equivalent to simple rotations and therefore we do not know if it can really be implemented experimentally. 

\subsection{$\pi$-pulse effect in the strongly synchronized regime}

In order to illustrate how the $\pi$-pulse perturbs the spin self-rephasing mechanism we present some typical simulation results obtained in a regime of strong synchronization (i.e. when $\omega_{ex} \gg \Delta_0$) which corresponds to the large density regime of the first experiment (Fig.1 of the main text). To simplify the discussion we take a collision rate $\gamma_c=0$ (i.e. no lateral collisions) such that the spin dynamics is now determined by only two time scales: the dephasing time $T_0=2\pi/\Delta_0$ and the exchange period $T_{ex}=2\pi/\omega_{ex}$. In the strongly synchronized regime ($T_{ex}\ll T_0$) and in absence of any $\pi$-pulse, the time dependent contrast $C(t)=|\bm{\bar{S}}(t)|$ presents periodic maxima at \textit{integer} times of the exchange period ($t=pT_{ex}$) and minima at \textit{half-integer} times ($t=(p+1/2)T_{ex}$). We can now introduce a $\pi$-pulse at a time $t_\pi$ and examine how the contrast is changed up to a time $t=2t_\pi$ at which we expect a spin-echo in absence of spin self-rephasing.
We consider two extreme cases: (i) the $\pi$-pulse occurs at a time $t_\pi=pT_{ex}$ which corresponds to a maximum of the contrast or (ii) the $\pi$-pulse occurs at a time $t_\pi=(p+1/2)T_{ex}$ which corresponds to a minimum of the contrast. More quantitatively, case (i) is implemented by taking $\Delta_0/2\pi=0.8$ Hz, $\omega_{ex}/2\pi=4$ Hz and $t_\pi=250$ ms such that $t_\pi=T_{ex}$. To obtain case (ii), instead of changing the value $t_\pi$ we choose to change the exchange rate to $\omega_{ex}/2\pi=6$ Hz such that now $t_\pi=\frac{3}{2} T_{ex}$.

The simulation results for case (i) are shown on figures Fig.1.a (no $\pi$-pulse) and Fig.1.b (integer $\pi$-pulse). 
Correspondingly, the results for case (ii) are shown on figures Fig.1.c (no $\pi$-pulse) and Fig.1.d (half-integer $\pi$-pulse). 
Let's consider first the black curves which represent the time dependent contrast $C(t)$. Comparing Fig.1.a and Fig.1.b, 
it seems that the integer $\pi$-pulse is a small perturbation only that slightly amplifies the variation of the contrast. 
On the other hand, when comparing Fig.1.c and Fig.1.d, it is immediate to see that the half-integer $\pi$-pulse constitutes 
a much stronger perturbation. More precisely, in this situation, the $\pi$-pulse strongly amplifies the contrast variation.  
For instance the minima become much deeper. Moreover both the positions of minima and maxima are strongly shifted in time.

In order to gain a deeper understanding on the origin of the contrast variation induced by the $\pi$-pulse, 
it appears instructive to look at the time dependent dynamics of the component $S_w(t)$, separately for cold (blue curves) and hot (red curves) atoms. 
More precisely, the blue curves are obtained from defining $S_w ^{\rm{cold}}(t)=\int _0^{E_{\rm{half}}} dE'\frac{E'^2}{2}e^{-E'}S_w(E',t)$ 
and the red curves from defining $S_w ^{\rm{hot}}(t)=\int _{E_{\rm{half}}} ^{\infty} dE'\frac{E'^2}{2}e^{-E'}S_w(E',t)$, where the energy scale $E_{\rm{half}}$ 
(obtained from $\int _0^{E_{\rm{half}}} dE'\frac{E'^2}{2}e^{-E'}=1/2$) separates the gas into two equally populated classes of hot and cold atoms. 
For the considered initial conditions, the hot and cold components verify the identity  $S_w ^{\rm{cold}}(t)=-S_w ^{\rm{hot}}(t)$ such that their sum vanishes at any time $t$.
Looking at figures Fig.1.(a,b,c,d), we observe that before the $\pi$-pulse ($t<t_\pi$) the hot component $S_w ^{\rm{hot}}(t)$ is positive definite whereas the cold one
$S_w ^{\rm{cold}}(t)$ is negative definite. In addition, for $t<t_\pi$ the hot and cold components of $S_w(t)$ oscillate periodically such that the minima of 
the contrast $C(t)$ coincide with the maxima of their modulus whereas the maxima of $C(t)$ are located at the minima of their modulus $|S_w ^{\rm{hot}}(t)|=|S_w ^{\rm{cold}}(t)|\approx 0$. 
Consider now the situation on figures Fig.1.(b,d) after the $\pi$-pulse ($t>t_\pi$). Since $S_w ^{\rm{cold}}(t)=-S_w ^{\rm{hot}}(t)$,
it is sufficient to analyze the behavior of the hot component $S_w ^{\rm{hot}}(t)$ which corresponds to the red curves. For $t>t_\pi$, 
we observe that $S_w ^{\rm{hot}}(t)$ still oscillates periodically but now also exhibits changes of sign. Interestingly,
the time averaged value of $S_w ^{\rm{hot}}(t)$ stays the same as before the $\pi$-pulse so that the variation amplitude of $S_w ^{\rm{hot}}(t)$ has necessarily increased. 
In fact, the minimal (negative) value of $S_w ^{\rm{hot}}(t)$ appears to exactly corresponds to minus the value of $S_w ^{\rm{hot}}(t=t_\pi)$ just before applying the pulse. 
As a consequence, for a half-integer $\pi$-pulse, the amplitude of variation of $S_w ^{\rm{hot}}(t)$ is expected to have been multiplied by a factor three (figure Fig.1.d). 
Furthermore, we note that the maxima of $S_w ^{\rm{hot}}(t)$ still coincide with the minima of the contrast $C(t)$ whereas the maxima of $C(t)$ now coincide 
with the new (negative) minima of $S_w ^{\rm{hot}}(t)$. The latter property constitutes the most surprising feature of the effect of the $\pi$-pulse.
It also explains the shift in position of the maxima of $C(t)$ which are now expected to occurs at time $t \approx t_\pi + p T_{ex}$.
To conclude, we note that the qualitative picture developed in this last paragraph is largely at the origin of the two macrospins schematic model detailed in the main text of the paper.

\begin{figure}[!ht]
{\includegraphics[width=15.cm]{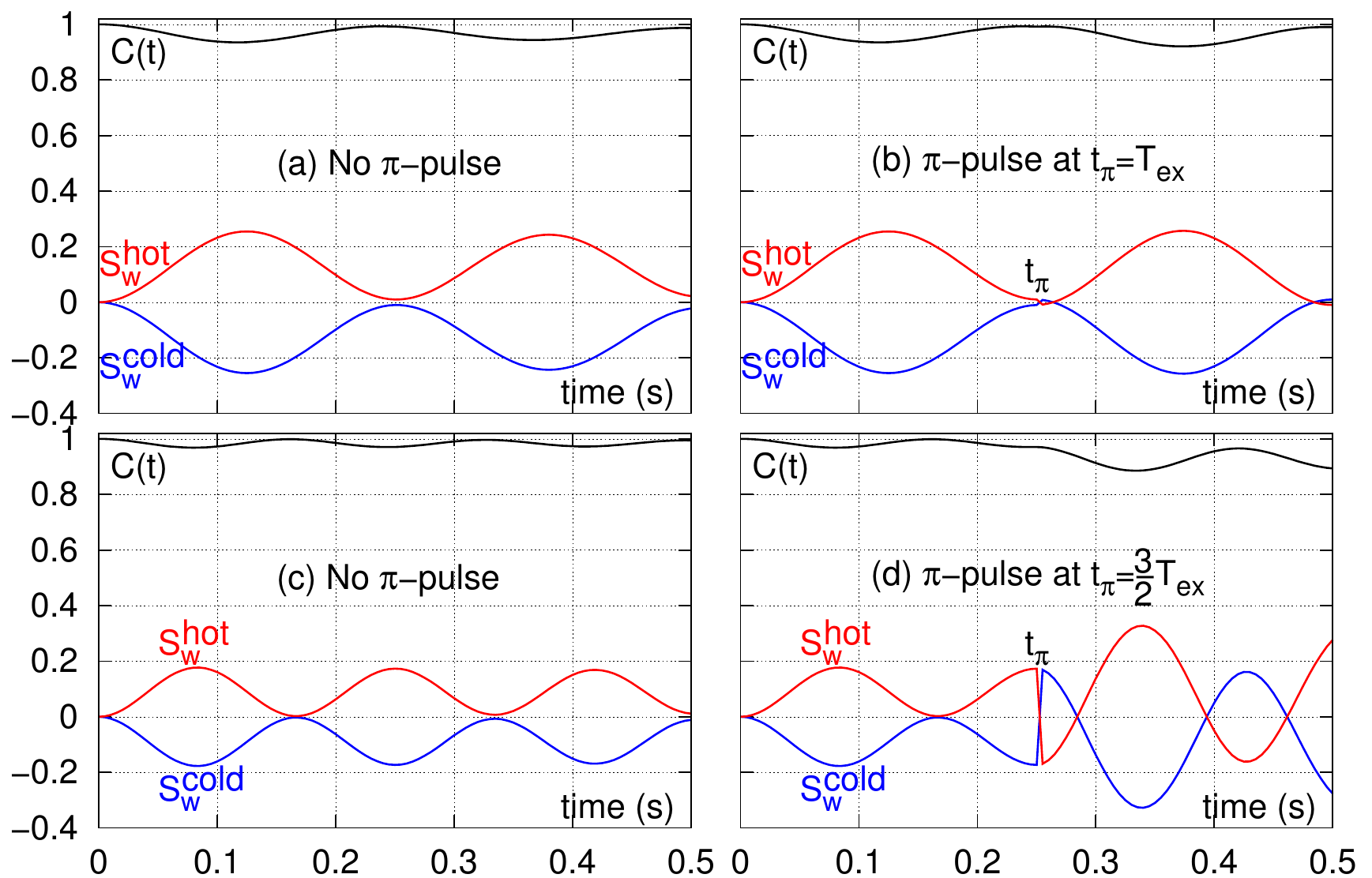}}
\caption{(Color online). The black curves represent the time dependent contrast $C(t)=|\bm{\bar{S}}(t)|$. 
The red curves represent the $w$ component $S_w ^{\rm{hot}}(t)$ of the hot atoms.
The blue curves represent the $w$ component $S_w ^{\rm{cold}}(t)$ of the cold atoms. The hot and cold components 
verify the identity  $S_w ^{\rm{cold}}(t)=-S_w ^{\rm{hot}}(t)$ such that their sum vanishes at any time $t$ (green zero line).
For all figures $\Delta_0/2\pi=0.8$ Hz and $\gamma_c=0$. Figs 1.(a,b) $\omega_{ex}/2\pi=4$ Hz. Figs 1.(c,d) $\omega_{ex}/2\pi=6$ Hz.
(a) No $\pi$-pulse. (b) $\pi$-pulse at $t_\pi=0.25$. The $\pi$-pulse occurs at integer exchange time: $t_\pi= T_{ex}$. (c) No-$\pi$-pulse.
(d) $\pi$-pulse at $t_\pi=0.25$ s. The $\pi$-pulse occurs at half-integer exchange time $t_\pi= \frac{3}{2} T_{ex}$.}
\end{figure}

\section{Fitting parameters for the second experiment}

The equations of motion (\ref{skees}) contains three experimentally tunable parameters: the inhomogeneity $\Delta_0$, the ISRE rate $\omega_{ex}$ and the lateral collision rate $\gamma_c$. As explained in the main text, within the considered experimental conditions, their predicted values are respectively $\Delta_0/2\pi\approx (0.3+0.7 \bar n)$ Hz, $\omega_{ex}/2\pi\approx 7.5 \bar n$ Hz and $\gamma_c\approx 2.4 \bar n$ s$^{-1}$ as function of the mean atom density $\bar n$ in unit $10^{12}$ at/cm$^3$. In practice, for each density we consider $\Delta_0,\omega_{ex}$ and $\gamma_c$ as fitting parameters in order to best fit the experimental data. As typical example, figures Fig.2.a,b,c illustrate the values found for $\Delta_0,\omega_{ex}$ and $\gamma_c$ in the case of the second experiment (Fig.3.a,b,c,d of the main text), where data obtained with or without $\pi$-pulse are fitted independently. The straight lines on these figures suggest that the best fits are obtained for $\Delta_0/2\pi\approx (0.8\pm 0.1)$ Hz, $\omega_{ex}/2\pi\approx (2.2\pm.1) \bar n$  Hz and $\gamma_c\approx (.53\pm .04) \bar n$ s$^{-1}$. The fitted values for $\Delta_0,\omega_{ex},\gamma_c$ show the same tendency as in previous experiments \cite{Deutsch_&al_2010} and \cite{Kleine_&al_2011}. In particular, we remind that the necessity to take smaller values for $\omega_{ex},\gamma_c$ can be attributed to the fact that these two contributions are taken in an infinite range approximation in eq.(\ref{skees}); this has a tendency to overestimate their effects. We note however that, in all experiments, the fitted values for $\Delta_0$ appear almost density independent in stricking contrast to the prediction. The latter feature still remains to be explained.

\begin{figure}[!ht]
\begin{center}
\begin{tabular}{ccc}

{\includegraphics[width=0.32\textwidth]{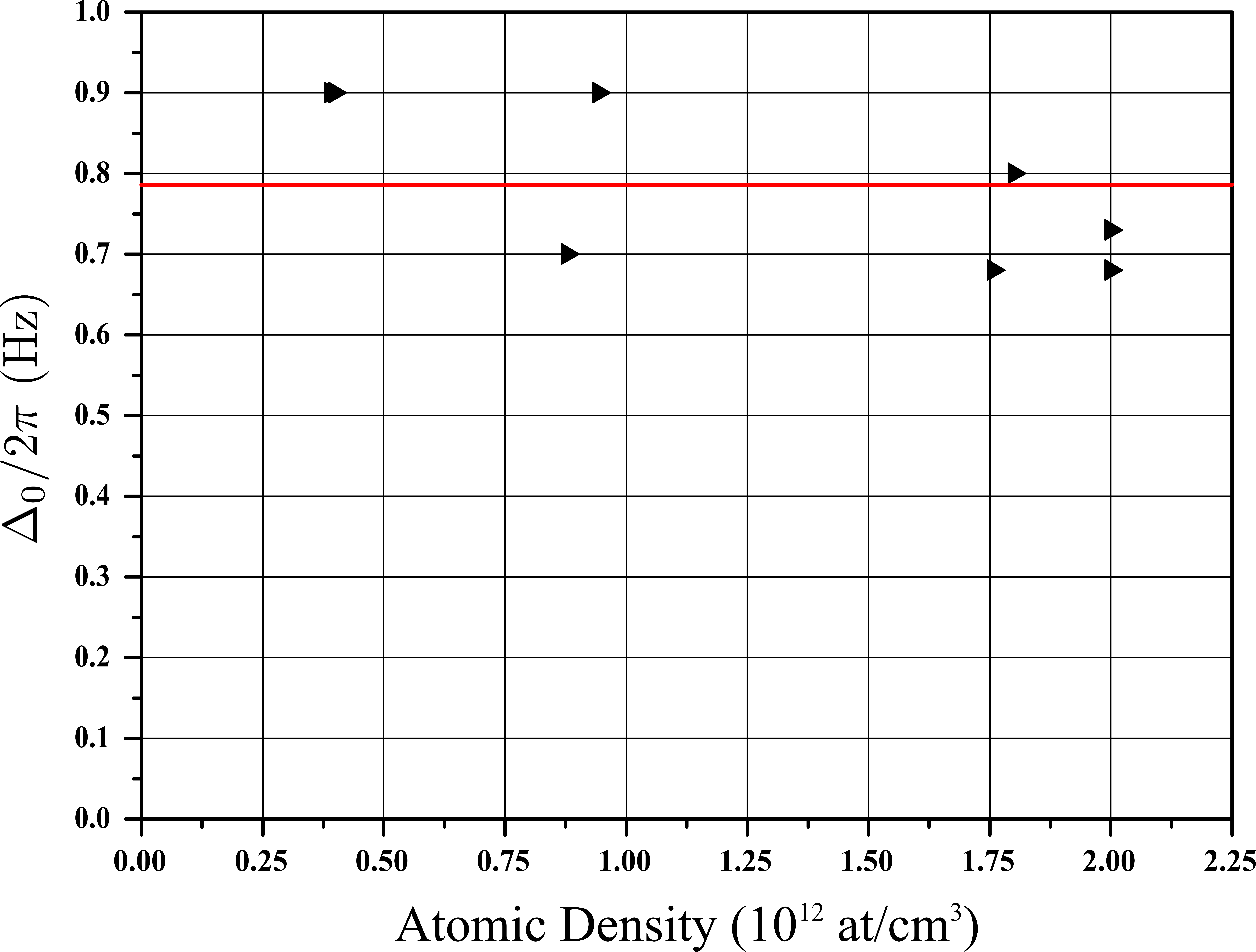}}&
{\includegraphics[width=0.32\textwidth]{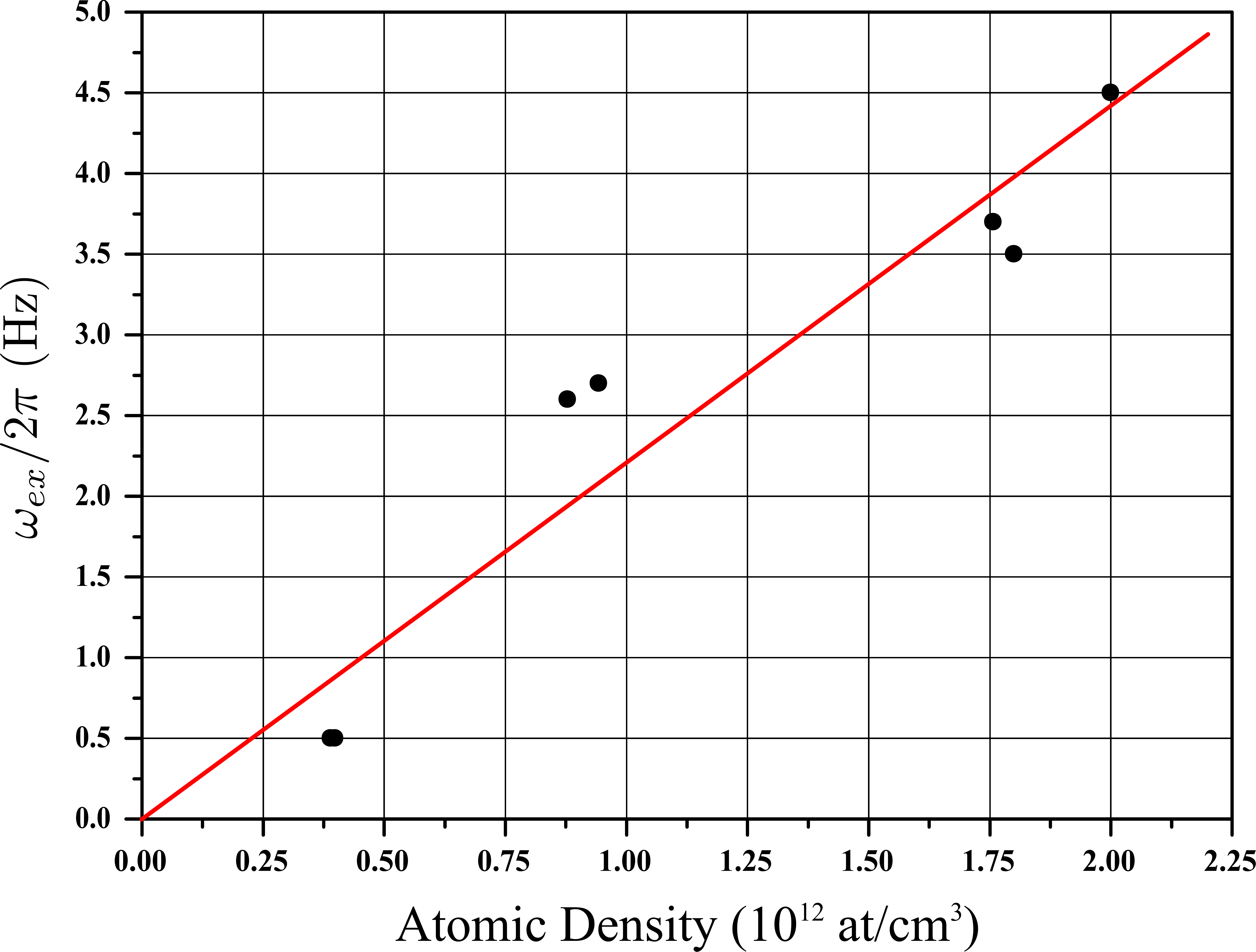}}&
{\includegraphics[width=0.32\textwidth]{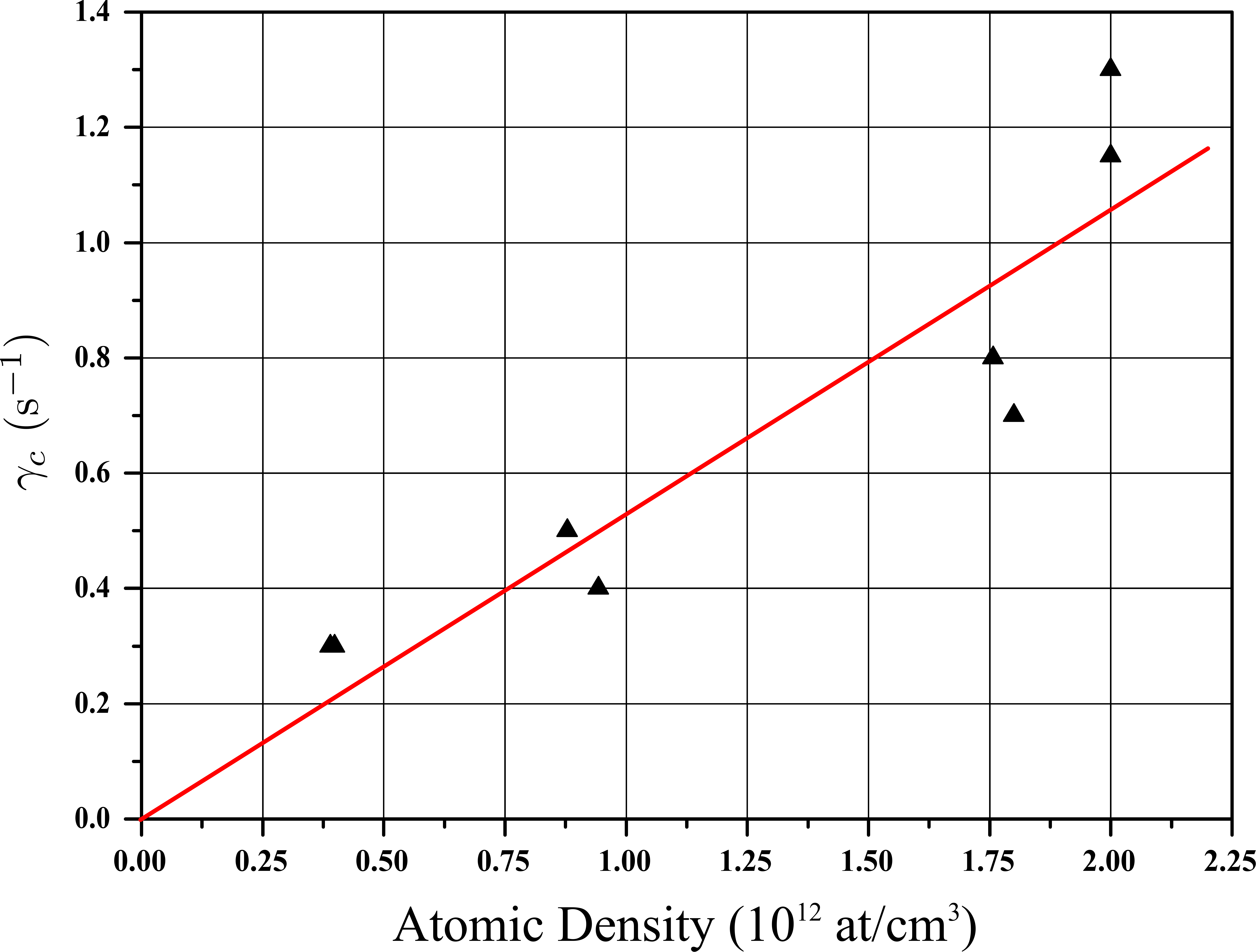}}
\\
 \large $(a)$&\large$(b)$&\large$(c)$
 \end{tabular}
\caption{(Color online). Fitting parameters corresponding to the second set of experiments of the main text: (a) the inhomogeneity $\Delta_0$, (b) the ISRE rate $\omega_{ex}$, (c) the collision rate $\gamma_c$. The straight lines on these figures  suggest that the best fits are obtained for 
$\Delta_0/2\pi\approx (0.8\pm 0.1)$ Hz , $\omega_{ex}/2\pi\approx (2.2\pm.1) \bar n$  Hz and $\gamma_c\approx (.53\pm .04) \bar n$ s$^{-1}$ which are to be compared with predicted values $\Delta_0/2\pi\approx (0.3+0.7 \bar n)$ Hz , $\omega_{ex}/2\pi\approx 7.5 \bar n$  Hz and $\gamma_c\approx 2.4 \bar n$ s$^{-1}$, with $\bar n$ the mean atom density in $10^{12}$ at/cm$^3$ unit.}
\label{figfits}
\end{center}
\end{figure}

\bibliographystyle{apsrev4-1}
\bibliography{BibliographieS}